\documentclass[10pt,conference]{IEEEtran}

\usepackage[utf8]{inputenc}
\usepackage[T1]{fontenc}
\usepackage{amsmath}
\usepackage{booktabs}
\usepackage{array}
\usepackage{graphicx}
\usepackage{xcolor}
\usepackage{tikz}
\usetikzlibrary{arrows.meta,positioning,calc}
\usepackage{url}
\usepackage[hidelinks]{hyperref}
\usepackage[capitalise,noabbrev]{cleveref}

\newcommand{\tool}{\textsc{SmartLens}}
\newcommand{\repo}[1]{\texttt{#1}}
\newcolumntype{L}[1]{>{\raggedright\arraybackslash}p{#1}}

\begin{document}

\title{\tool{}: Turning AI Agent Changes into Learning, Review, and Risk Signals in the IDE}

\author{\IEEEauthorblockN{Vahid Etemadi}
\IEEEauthorblockA{Independent Researcher\\
Shiraz, Iran\\
vetemadi87@gmail.com}
\and
\IEEEauthorblockN{Gregorio Robles}
\IEEEauthorblockA{Universidad Rey Juan Carlos\\
Madrid, Spain\\
grex@gsyc.urjc.es}}

\maketitle

\begin{abstract}
Developers increasingly act as \emph{managers of AI agents}: they hand tasks to AI coding agents and then have
to understand, verify, ship, and learn from what the agents produce. IDEs and recent tools can show
which committed lines were written by AI, but they do not attribute the pending changes of external agents, and
they do not use attribution to help the developer review, learn, or judge risk. We present \tool{}, an IntelliJ IDEA plugin that attributes changes to an AI agent, the
developer, or both. It combines this attribution with the Git history to support three goals: (i) \emph{Learn} compares the skills (APIs, language features, annotations) that agents used with the skills
the developer has used in their own commits; (ii) \emph{Understand} ranks pending changes by review
priority; and (iii) \emph{Risk} shows hotspots: files that change often, are mostly written by AI, and that the
developer barely knows. In a preliminary evaluation, \tool{} attributed 8 of 12 scripted IDE scenarios correctly;
all failures were in (expected) blind spots. On 8 Open Source Software repositories with agent co-authored commits, agents
introduced up to 253 skills that do not appear in the main developer's own commits. The Risk ranking was as
good as or better than plain churn at predicting later bug fixes, but part of this gain is due to recency.
\end{abstract}

\begin{IEEEkeywords}
AI coding agents, IDE, program comprehension, code review, change attribution, cognitive debt
\end{IEEEkeywords}

\section{Introduction}

AI coding agents do more than complete lines: they edit many files, run commands, and commit on the
developer's behalf~\cite{yang2024sweagent,jimenez2024swebench}, and they are already common in Open Source Software (OSS) 
projects~\cite{robbes2026agentic,li2026aidev}. As a result, developers of all levels of experience increasingly
work as \emph{managers} of agents~\cite{robles2027wanted}: they describe a task, delegate it, and get back a
set of changed files. As in any delegation, the responsibility stays with the one who delegates. The developer
still has to \emph{understand}, \emph{verify}, \emph{ship}, \emph{decide}, and ideally \emph{learn} from what was
changed, often having seen little more than the agent's summary and a diff.

This leads to three problems. First, checking agent output takes time: developers spend a lot of it
reading and verifying generated code~\cite{mozannar2024reading,liang2024survey}, and may even end up slower
than without AI~\cite{becker2025metr}. Second, delegating creates \emph{cognitive
debt}~\cite{kosmyna2025brain}: the developer no longer knows the details of the delegated work. Robles et
al.~\cite{robles2027wanted} argue that this debt is not a flaw of the tools but is inherent to any delegation,
just as no manager knows every detail of the work of their team. It becomes harmful when the developer can no
longer recover the knowledge needed to change the work, so it has to be managed, e.g., by revisiting delegated
work before building on it. Moreover, developers stop practicing what they delegate~\cite{risko2016offloading}, a concern raised for novices learning with code
generators~\cite{kazemitabaar2023novices,prather2023robots}, and a recent experiment found that developers who
delegated to AI learned less about a new library~\cite{shen2026skill}. Third, code that changes often and that
nobody on the team really owns is a known predictor of defects~\cite{nagappan2005churn,bird2011ownership};
agents can increase both churn and the loss of ownership.

Current IDEs offer only partial help here. Git blame shows who committed each line, and recent tools also
record which committed lines were written by AI~\cite{jetbrains2026aianalytics,gitai2026}. However, these tools
only cover committed code, and some only cover agents that run inside the IDE. Before a commit, the IDE shows a line written
by an agent in the same way as a line typed by the developer. Moreover, these tools use attribution for blame or for usage
dashboards, not to support the developer while reviewing. This is a missed opportunity: unlike the work of a
human employee, every action of an agent can be observed and logged at little cost~\cite{robles2027wanted}. We argue that \emph{change attribution}, i.e., knowing who produced a change, is what an IDE needs to turn agent
activity into useful signals. In terms of cognitive debt, attribution
tells the developer which parts of the code they no longer know, where it is worth paying the debt back, and what
they could learn from the agent. We call this last practice \emph{reverse learning}: instead of humans
teaching the machine, the developer learns from what the agent did.

The contributions of this paper are:
\begin{itemize}
  \item A lightweight approach, at the IDE level, that attributes code changes to an AI agent, the developer, or both.
        It combines editor and file-system events for uncommitted changes with commit trailers for the history,
        and can use an agent hook log for more precision (\cref{sec:attribution}).
  \item Three views built on attribution and Git history: \emph{Learn}, \emph{Understand}, and \emph{Risk},
        each based on a simple and transparent scoring model (\cref{sec:views}).
  \item \tool{}, an implementation for IntelliJ IDEA (\cref{sec:implementation}), and a preliminary evaluation
        with scripted IDE scenarios and eight OSS repositories (\cref{sec:evaluation}).
\end{itemize}

\section{Developer Goals When Managing Agents}
\label{sec:goals}

Besides getting the task done, a developer who manages agents has three goals. They correspond to three ways
of managing the cognitive debt created by delegation.

\textbf{G1: Learn new things from the agent.} Agents often use APIs, idioms, or language features that the
developer has not used before. If they are pointed out at the right time, these are chances to learn.
If not, they are skipped in a quick review and end up as code the developer cannot maintain.

\textbf{G2: Verify with little effort.} Not every change needs the same attention. A large change made only
by an agent in a file the developer barely knows needs more attention than a small edit
the developer typed themselves. Reviewing changes in this order helps the developer spend their limited
attention where it is most needed~\cite{bacchelli2013review}, and pay back the debt before building on the change.

\textbf{G3: Keep the most-changed parts maintainable.} Files that have changed many times, have many authors,
and were mostly written by agents are likely places for regressions. Developers should know where these
hotspots, with high and growing debt, are before they accept more agent changes there.

All three goals need two pieces of information that IDEs do not fully track today: \emph{who} changed a file
(agent, developer, or both), including changes not yet committed, and \emph{what the developer already knows}, which we estimate from their own
Git history~\cite{mockus2002expertise,fritz2010dok}.

\section{Approach}
\label{sec:approach}

\begin{figure}[t]
\centering
\resizebox{\columnwidth}{!}{%
\begin{tikzpicture}[
  font=\footnotesize,
  box/.style={draw, rounded corners=2pt, align=center, minimum height=0.8cm, inner sep=3pt, fill=white},
  src/.style={box, fill=blue!6},
  view/.style={box, fill=green!8, minimum width=1.9cm},
  arr/.style={-{Stealth[length=2mm]}, thick}
]
  \node[src] (editor) {Editor\\typing};
  \node[src, right=0.25cm of editor] (vfs) {External disk\\changes};
  \node[src, right=0.25cm of vfs] (hook) {Agent\\hook log};
  \node[src, right=0.25cm of hook] (git) {Git history\\\& trailers};

  \node[box, below=0.7cm of $(vfs)!0.5!(hook)$, minimum width=6.2cm] (attr)
       {Attribution: \textsc{Agent} / \textsc{Mixed} / \textsc{Developer}};
  \node[box, below=0.45cm of attr, minimum width=6.2cm] (model)
       {Developer vocabulary \& familiarity, file churn};

  \node[view, below=0.7cm of model] (understand) {Understand};
  \node[view, left=0.3cm of understand] (learn) {Learn};
  \node[view, right=0.3cm of understand] (risk) {Risk};

  \draw[arr] (editor) -- (attr);
  \draw[arr] (vfs) -- (attr);
  \draw[arr] (hook) -- (attr);
  \draw[arr] (git.east) -- ++(0.3,0) |- (model.east);
  \draw[arr] (attr) -- (model);
  \draw[arr] (model) -- (learn);
  \draw[arr] (model) -- (understand);
  \draw[arr] (model) -- (risk);
\end{tikzpicture}%
}
\caption{Overview of \tool{}. Change signals are combined into an attribution of pending changes, which,
together with a model of the developer derived from Git history, feeds three views.}
\label{fig:overview}
\end{figure}
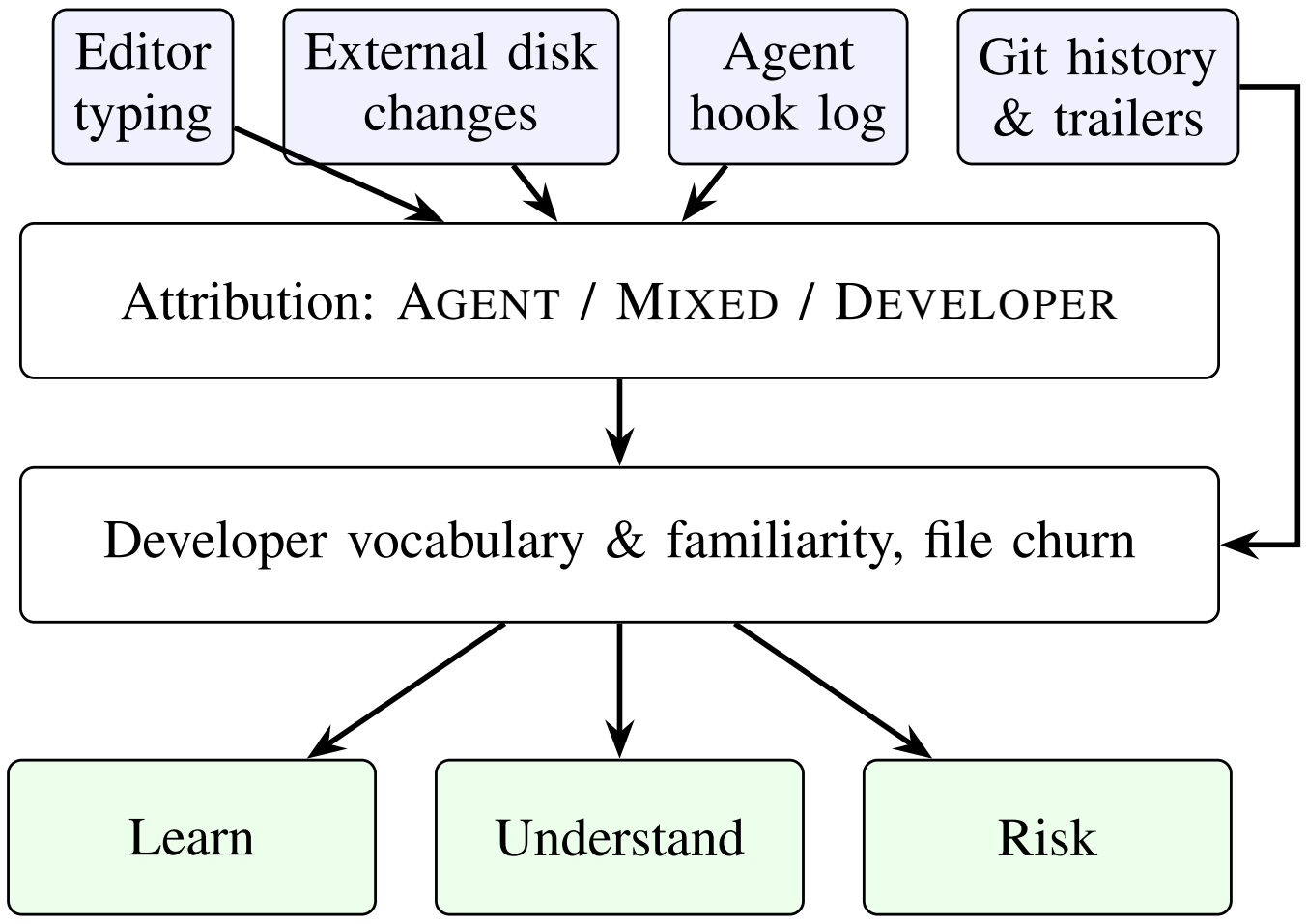

\cref{fig:overview} gives an overview. \tool{} collects attribution signals while the developer
and their agents work, and computes the three views on request from the attribution and the Git history.
We use the term \emph{pending changes} for uncommitted modifications, including new untracked files.

\subsection{Change Attribution}
\label{sec:attribution}

No single signal can reliably separate agent edits from developer edits, so \tool{} combines the signals in
\cref{tab:signals}.

\begin{table}[t]
\centering
\caption{Attribution signals used by \tool{}.}
\label{tab:signals}
\footnotesize
\begin{tabular}{@{}L{2.3cm}L{1.4cm}L{3.8cm}@{}}
\toprule
\textbf{Signal} & \textbf{Attributed to} & \textbf{Caveat} \\
\midrule
File changed on disk outside the IDE & Agent & Also fires for Git operations and external formatters; Git operations are filtered out. \\
Edit typed in the IDE editor & Developer & In-IDE assistants that write through the editor look like the developer. \\
Agent hook log entry & Agent & Most precise; requires a one-line hook per agent. \\
Commit author or \texttt{Co-\allowbreak authored-\allowbreak by} trailer matches an agent & Agent (history) & Depends on commit conventions; attributes whole commits. \\
\bottomrule
\end{tabular}
\end{table}

\emph{External changes.} CLI agents such as Claude Code, Codex, or Aider write files directly to disk. The
IDE notices these changes when it refreshes its virtual file system, and \tool{} records them as agent edits.
Git itself also changes files on disk (checkout, rebase, merge), which would be wrongly attributed to an agent.
\tool{} detects Git operations through changes to Git's state files (e.g., \texttt{HEAD}) and through repository
events, and ignores file changes for a short quiet period (5\,s by default) after them.

\emph{Developer edits.} Edits made in the IDE editor are recorded as developer edits. When the IDE reloads a
document because of an external change, this reload is not counted, so the agent's edit is not counted twice.

\emph{Agent hook log.} Many agents support lifecycle hooks. A one-line hook adds a JSON record
(\texttt{file}, \texttt{agent}, timestamp) to a log in the project after every edit. This is the most precise
signal, and it also covers agents that edit through the IDE.

\emph{Git history.} For committed work, \tool{} marks a commit as an agent commit if its author or a
\texttt{Co-authored-by} trailer matches a configurable list of agent identities.

For every file with pending changes, \tool{} counts agent edits $a$ and developer edits $d$ since the last
commit and classifies the file as \textsc{Agent} ($a>0, d=0$), \textsc{Mixed} ($a>0, d>0$),
\textsc{Developer} ($a=0, d>0$), or \textsc{Unknown}. Each class has an \emph{origin score}
$o \in \{1.0, 0.6, 0.0, 0.3\}$, respectively: a file written only by an agent needs more attention
than one the developer also edited. Attribution records are removed once a file has no pending changes.

\subsection{Developer Model}
\label{sec:model}

\tool{} identifies the developer by their Git e-mail address (and optional aliases) and analyzes the latest $N$
commits (300 by default). For each file $f$, it computes the number of commits $c_f$, the number of authors $u_f$,
the share of added lines that come from agent commits ($s_f$), and the developer's \emph{familiarity} $k_f$, i.e., the
share of added lines the developer wrote themselves. This is a simple approximation of the
authorship part of degree-of-knowledge models~\cite{fritz2010dok,fritz2007activity}.

\subsection{The Three Views}
\label{sec:views}

\textbf{Learn (G1).} \tool{} extracts \emph{features} from added lines: imported APIs, annotations, and
language features such as records, sealed types, pattern matching, streams, or try-with-resources. The
extraction is lexical, because it has to work on diff fragments that cannot be parsed. The developer's
\emph{vocabulary} $V_{dev}$ is the set of features found in lines added by their own commits and by their own
pending changes. The agent's usage $V_{agent}$ is collected from agent commits and agent pending
changes. \tool{} reports $V_{agent} \setminus V_{dev}$, without imports of the project's own packages,
ordered by how often the agent used each feature, with a short explanation and a link to an example. In this way,
delegated work becomes a concrete, personal list of things to learn. It encourages the developer to engage
with the new material, which supports learning~\cite{bjork1994desirable}, instead of fully offloading it.

\textbf{Understand (G2).} Pending changes are ranked by a review priority
\begin{equation}
  p = 100 \cdot \bigl(0.5\,o + 0.3\,\sigma + 0.2\,(1-k_f)\bigr),
\end{equation}
where $o$ is the origin score and $\sigma = \min\bigl(1, \log(1+\Delta)/\log(301)\bigr)$ normalizes the
number of changed lines $\Delta$. Equation~(1) is our own heuristic, not an established model, but each term
is based on prior work. The origin term reflects that AI-generated code needs careful
verification~\cite{perry2023insecure,mozannar2024reading}; the size term, that defects in larger patches are more
likely to be missed in review~\cite{kononenko2015review}; and the familiarity term, that code a developer
knows little about is harder to judge and more defect-prone~\cite{fritz2010dok,bird2011ownership}. The order also
matters because reviewers find more defects in the files they see first~\cite{fregnan2022first}. Opening an
entry shows the IDE's diff, so the developer reviews the riskiest changes first and can stop when only their
own small edits are left.

\textbf{Risk (G3).} Following hotspot analysis~\cite{tornhill2015crime} and churn-based defect
prediction~\cite{nagappan2005churn,hassan2009entropy}, \tool{} gives each file the score
\begin{equation}
  r = 100 \cdot \chi_f \cdot \bigl(0.4 + 0.3\,s_f + 0.3\,(1-k_f)\bigr) \cdot \alpha_f,
\end{equation}
where $\chi_f = \log(1+c_f)/\log(1+c_{max})$ is normalized churn and
$\alpha_f = 0.85 + 0.15\min\bigl(1, (u_f-1)/4\bigr)$ accounts for scattered ownership~\cite{bird2011ownership}.
Churn is still the main factor: a file that changes all the time is a hotspot even if the developer wrote
it, but it ranks higher if agents wrote it and the developer does not know it. Dotfiles and Markdown
files are ignored, and the view marks hotspots that currently have pending agent changes.

The weights are simple and documented; we did not tune them on the data of
\cref{sec:evaluation}.

\section{Implementation}
\label{sec:implementation}

\begin{figure*}[t]
\centering
\includegraphics[width=\textwidth]{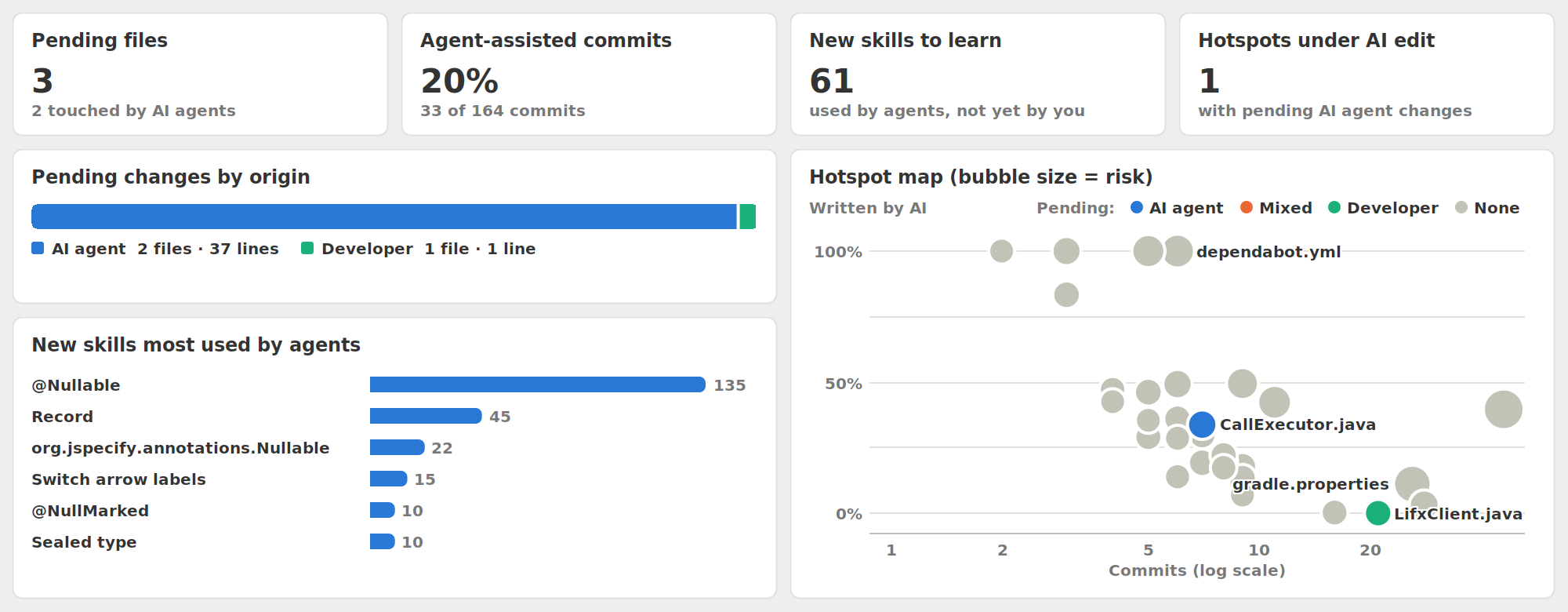}\\[2pt]
{\footnotesize (a) Overview dashboard: headline numbers, pending changes by origin, the new skills agents used most, and
the hotspot map (commits on a log scale vs.\ share written by AI; bubble area = risk; colored bubbles have pending
changes, so \repo{CallExecutor.java} stands out as a hotspot under agent edit).}\\[4pt]
\includegraphics[width=\textwidth]{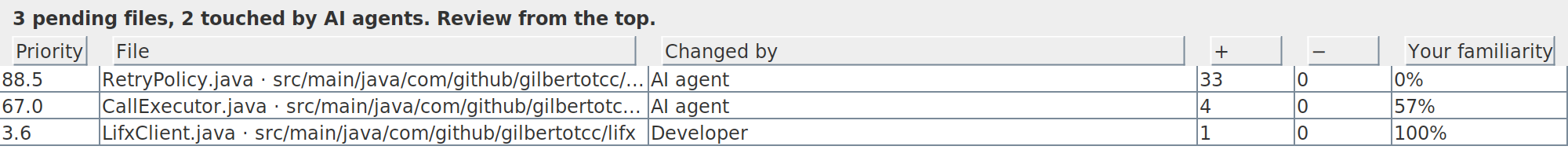}\\[2pt]
{\footnotesize (b) Understand table view: pending changes ranked by review priority; Learn and Risk have equivalent tables.}
\caption{The \tool{} tool window, rendered from the plugin running headless on
\repo{gilbertotcc/lifx-client} (164 commits, 33 agent-assisted), after a scripted agent session that added
\repo{RetryPolicy.java} and changed \repo{CallExecutor.java} while the developer edited \repo{LifxClient.java}.
Learn therefore reports one skill more than \cref{tab:mining}, which covers committed history only.}
\label{fig:screenshot}
\end{figure*}

\tool{} is implemented in Java as an IntelliJ Platform plugin (IntelliJ IDEA 2025.1 and later) on top of the
bundled Git integration, with about 2.3k lines of code in 49 classes. It is opened via
\emph{Tools\,|\,SmartLens}, which shows a tool window with an Overview dashboard and one table tab per view
(\cref{fig:screenshot}).
Attribution is collected by three listeners registered in the plugin descriptor: a virtual-file-system listener for
external changes, a document listener for editor edits, and a Git repository listener for VCS operations.
Attribution data is stored in the IDE's per-user workspace file, so it stays on the developer's machine.
The analysis runs as a background task that can be cancelled. It runs \texttt{git log}, \texttt{git diff}, and
\texttt{git ls-files} once per repository and sends the result to the tool window over the IDE's message
bus. The scoring code does not depend on the IDE and has unit tests; an end-to-end test runs the plugin in a headless
IDE on a real Git repository. \tool{} needs no network access and no LLM. The plugin, the study scripts, and
the data will be released as open source.\footnote{\url{https://github.com/vahidetemadi/smartlens}}

\section{Preliminary Evaluation}
\label{sec:evaluation}

We address two research questions. \textbf{RQ1}: How reliable is the attribution, and where does it fail?
\textbf{RQ2}: What do the Learn and Risk views report on real repositories with agent activity?

\subsection{RQ1: Attribution in Scripted IDE Scenarios}

We wrote 12 scenarios that run in a headless IntelliJ instance with the plugin installed, on a real
Git repository. Each scenario changes its own file, so the correct answer is known. We simulate agents by
writing files to disk from outside the IDE, the developer by editing documents through the editor API, and
Git operations by running \texttt{git}. Besides the normal cases, we included on purpose the blind spots
from \cref{tab:signals}. We ran all scenarios with the default 5\,s Git quiet period and, as an ablation,
without it.

\begin{table}[t]
\centering
\caption{Attribution scenarios: expected vs.\ observed origin (A = \textsc{Agent}, D = \textsc{Developer},
M = \textsc{Mixed}, -- = not reported), with and without the Git quiet period.}
\label{tab:scenarios}
\footnotesize
\begin{tabular}{@{}L{4.55cm}ccc@{}}
\toprule
\textbf{Scenario} & \textbf{Exp.} & \textbf{5\,s} & \textbf{0\,s} \\
\midrule
Agent edits a file that is not open & A & A & A \\
Agent edits a file open in the editor & A & A & A \\
Agent creates a new file & A & A & A \\
Developer types in the editor & D & D & D \\
Developer edits, then agent edits & M & M & M \\
Agent edits, developer edits 4\,s later & M & M & M \\
Agent edits, developer edits within 1\,s & M & \textbf{A} & \textbf{A} \\
In-IDE assistant edits, no hook & A & \textbf{D} & \textbf{D} \\
In-IDE assistant edits, with hook entry & A & \textbf{M} & \textbf{M} \\
Developer runs an external formatter & D & \textbf{A} & \textbf{A} \\
Agent edits a git-ignored file & -- & -- & -- \\
\texttt{git checkout}, then developer edits & D & D & \textbf{M} \\
\midrule
Correct & & 8/12 & 7/12 \\
\bottomrule
\end{tabular}
\end{table}

\cref{tab:scenarios} shows that \tool{} handles the common cases. This includes the tricky case in which the IDE
reloads an open document after an agent's change, which must not be counted as developer typing. All four
failures are blind spots we expected: developer typing within the 3\,s reload window after an agent
edit is lost; in-IDE assistants that write through the editor look like the developer (and with a hook entry
become \textsc{Mixed} instead of \textsc{Agent}); and external formatters look like agents. The ablation shows
that the Git quiet period is needed: without it, a checkout is taken for an agent edit.

\subsection{RQ2: Mining OSS Repositories}

We searched GitHub for commits with a \texttt{Co-Authored-By: Claude} trailer, kept Java and Kotlin
repositories, and selected eight that cover three situations: long human histories with recent agent use,
projects mostly written by agents from the start, and a control with almost no agent use (\cref{tab:mining}). For each repository, we ran \tool{}'s
history analysis (the same classes the plugin uses) on up to 2{,}000 non-merge commits. We took the most active
human author, with all their e-mail addresses, as the developer.

\begin{table}[t]
\centering
\caption{Repositories, agent activity, and Learn results. AI lines: share of added Java/Kotlin lines from agent
commits; $|V_{dev}|$: developer vocabulary; New: features used by agents but absent from $V_{dev}$.}
\label{tab:mining}
\scriptsize
\setlength{\tabcolsep}{3pt}
\begin{tabular}{@{}lrrrrr@{}}
\toprule
\textbf{Repository} & \textbf{Commits} & \textbf{Agent} & \textbf{AI lines} & $|V_{dev}|$ & \textbf{New} \\
\midrule
\repo{violations-gradle-plugin} & 317 & 16 (5\%) & 5\% & 56 & 3 \\
\repo{asciidoctorj-pdf} & 108 & 1 (1\%) & 0\% & 35 & 0 \\
\repo{lifx-client} & 164 & 33 (20\%) & 19\% & 256 & 60 \\
\repo{steamclog-android} & 89 & 19 (21\%) & 35\% & 82 & 25 \\
\repo{libseqera} & 370 & 155 (42\%) & 60\% & 202 & 145 \\
\repo{wisp} & 218 & 140 (64\%) & 95\% & 64 & 253 \\
\repo{kotlinstall} & 75 & 70 (93\%) & 100\% & 0 & 120 \\
\repo{xemantic-kotlin-script} & 43 & 42 (98\%) & 97\% & 12 & 161 \\
\bottomrule
\end{tabular}
\end{table}

\textbf{Learn.} Even where the developer has a long history of their own, agents introduced dozens of
features new to them, e.g., JSpecify nullness annotations, records, and arrow-form \texttt{switch} in
\repo{lifx-client}, or Kotlin coroutines and Room annotations in \repo{wisp}. In \repo{steamclog-android}
and \repo{kotlinstall}, the most frequent new items are testing APIs (JUnit assertions, \texttt{@Test}):
agents wrote tests where the developer had not. We found two issues. First, our first version reported the
project's own classes as ``skills''; we fixed this by excluding imports of packages declared in the project.
Second, in repositories written mostly by agents, almost every commit is co-authored. The vocabulary from the
repository is then nearly empty and every agent feature counts as new; in these cases, the vocabulary has to come
from the developer's other repositories.

\begin{table}[t]
\centering
\caption{Risk ranking evaluated against bug-fix commits: files ranked on the first 80\% of the history,
precision of the top 10 at predicting files touched by fix commits in the remaining 20\%.}
\label{tab:risk}
\scriptsize
\setlength{\tabcolsep}{2.2pt}
\begin{tabular}{@{}lrrrrrr@{}}
\toprule
\textbf{Repository} & \textbf{Files} & \textbf{Fixes} & \textbf{Base} & \textbf{Churn} & \textbf{Recent} & \textbf{\tool{}} \\
\midrule
\repo{violations-gradle-plugin} & 14 & 6 & .14 & .20 & .20 & .20 \\
\repo{asciidoctorj-pdf} & 24 & 2 & .25 & .30 & .50 & .30 \\
\repo{lifx-client} & 74 & 3 & .38 & .20 & .20 & .30 \\
\repo{steamclog-android} & 34 & 1 & .06 & .20 & .20 & .20 \\
\repo{libseqera} & 366 & 20 & .18 & .20 & .70 & .50 \\
\repo{wisp} & 110 & 9 & .06 & .10 & .20 & .30 \\
\bottomrule
\end{tabular}
\end{table}

\textbf{Risk.} We ranked files on the first 80\% of each history and checked which of the top 10 were changed
by fix commits (subject containing \emph{fix}, \emph{bug}, \emph{regression}, or \emph{revert}) in the remaining
20\% (\cref{tab:risk}). We compare against plain churn and against recent churn (commits in the last quarter
of the training period). We left out \repo{kotlinstall} and \repo{xemantic-kotlin-script} because their test
periods contain at most one fix. \tool{}'s ranking was never worse than plain churn, and it was better on three
repositories, where it moves up files with a high agent share (in \repo{wisp}, all top-10 files are more than
half written by AI). Compared with recent churn, however, the results are mixed: \tool{} is better on two repositories and
worse on two. Files written by agents have, by definition, been changed recently, so part of the gain over plain churn is
due to recency; we plan to add an explicit recency term. With one to 20 fix commits per
repository, these numbers are only indicative.

\subsection{Threats to Validity}

The scenarios are synthetic: they measure behavior on known cases, not accuracy in real use. The repositories
are a small convenience sample, found through a single trailer convention; agent commits without trailers are
counted as human. Familiarity uses authorship only, and fix commits are identified by keywords, a common but
noisy heuristic. Learn reports features, but not whether developers actually learn from them; this requires a
user study.

\section{Related Work}
\label{sec:related}

\textbf{Programmers and AI assistants.} Studies of code-generation tools report that developers find it hard to
understand and validate generated code~\cite{vaithilingam2022expectation,barke2023grounded}, that verification
takes a large share of their time~\cite{mozannar2024reading}, and that usability problems, such as output that is
hard to control, remain~\cite{liang2024survey}. Developers using assistants may also write less secure
code while feeling more confident~\cite{perry2023insecure}. In a randomized trial, experienced OSS
developers were slower with AI tools, although they believed they were faster~\cite{becker2025metr}.

\textbf{Coding agents in practice.} Autonomous agents~\cite{yang2024sweagent} increase the amount of such code.
Mining studies have used commit trailers and similar traces, as we do, to measure the adoption of agents on
GitHub~\cite{robbes2026agentic}, and have collected large datasets of agent-authored pull requests~\cite{li2026aidev}.
Most pull requests created with Claude Code are merged, but many need human revisions first~\cite{watanabe2025agentic}.
Drawing on management theories of delegation, Robles et al.\ frame these developers as
managers~\cite{robles2027wanted}. \tool{} does not generate code; it helps them handle what agents produced.

\textbf{Tools for AI attribution.} JetBrains' AI analytics report the share of committed lines written by
AI, but only for AI tools inside the IDE, not for agents in the terminal~\cite{jetbrains2026aianalytics}. Git AI, an OSS Git extension, records at commit time which lines
were written by an agent and shows them in an ``AI blame'' view~\cite{gitai2026}. Both stop at committed code and use attribution
for blame or metrics; \tool{} also covers pending changes and uses attribution for review, learning, and risk.

\textbf{Learning with AI.} Computing-education research discusses both the opportunities and the risk of
over-reliance when learners use code generators~\cite{kazemitabaar2023novices,prather2023robots}, which is related
to cognitive offloading~\cite{risko2016offloading}. Knowledge workers report less critical thinking effort when
they trust generative AI~\cite{lee2025critical}, and developers who delegated to AI scored lower on a test about a
library they had just used~\cite{shen2026skill}. \tool{} targets professional developers and derives
what to learn from the difference between the agent's output and the developer's own history.

\textbf{Expertise, ownership, and hotspots.} Expertise can be mined from version history~\cite{mockus2002expertise},
and degree-of-knowledge models combine authorship and interaction~\cite{fritz2010dok,fritz2007activity}.
Low ownership and high churn are associated with defects~\cite{bird2011ownership,nagappan2005churn,hassan2009entropy},
and hotspot analysis uses this in practice~\cite{tornhill2015crime}. We apply these ideas to development with
agents by separating agent authorship from human authorship.

\textbf{Comprehension and review.} Program comprehension is driven by hypotheses~\cite{brooks1983comprehension}, and
code review is mostly about understanding the change~\cite{bacchelli2013review}. The order in which files are shown
affects which defects reviewers find~\cite{fregnan2022first}. \tool{} shows the developer what they did not write
and in which order to read it.

\section{Limitations and Future Work}
\label{sec:future}

The attribution is based on a heuristic (\cref{tab:scenarios}): in-IDE assistants count as developer edits unless they use the
hook log, external formatters count as agent edits, developer typing right after an agent edit can be lost,
and co-authored commits are attributed as a whole. Familiarity uses only authorship, not reading activity, and
developers with several Git identities have to list their aliases. The feature extraction is lexical and designed
for Java and Kotlin. As next step, we plan a user study in which participants review change sets made by agents with
seeded defects, with and without \tool{}, and are tested one week later on the skills reported by Learn.
Other technical improvements are attribution at the method level using the IDE's code model, a developer
vocabulary that spans several repositories for projects written mostly by agents, a recency term in the Risk score,
and explanations and exercises for Learn items generated by an LLM.

Seeing developers as managers~\cite{robles2027wanted} suggests further directions. Linking each change to the
agent session that produced it (task, plan, decisions) would show not only \emph{who} changed the code but also
\emph{why}, keeping the knowledge recoverable. Since autonomy should depend on how reliable an agent has proven
on a kind of task, the mined history could show where agent changes were later fixed or reverted, and thus where
agents need closer supervision. Finally, team leads and maintainers who supervise people who in turn supervise
agents usually see only the final result; a review version of \tool{} could show them what was delegated and
what the author changed by hand.

\section{Conclusion}

When developers manage agents, they accept some cognitive debt, and the IDE should help them manage it. It should show not only what changed, but also who changed it, what the
developer can learn from it, and where it may break. \tool{} makes change attribution a core IDE
concept and builds three views on it. Our preliminary evaluation shows that simple attribution works for
common workflows and fails in predictable cases, that agents introduce many skills developers have not used, and
that agent authorship is a useful risk signal, although it is partly confounded with recency.

\section*{Acknowledgment}
We used an AI agent (Claude Opus 5.5) to help implement \tool{} and the study
scripts, run the studies, and draft this paper. The authors directed the work, reviewed and checked all code,
results, and text, and take full responsibility for the content.

\bibliographystyle{IEEEtran}
\bibliography{references}

\begin{thebibliography}{10}
\providecommand{\url}[1]{#1}
\csname url@samestyle\endcsname
\providecommand{\newblock}{\relax}
\providecommand{\bibinfo}[2]{#2}
\providecommand{\BIBentrySTDinterwordspacing}{\spaceskip=0pt\relax}
\providecommand{\BIBentryALTinterwordstretchfactor}{4}
\providecommand{\BIBentryALTinterwordspacing}{\spaceskip=\fontdimen2\font plus
\BIBentryALTinterwordstretchfactor\fontdimen3\font minus
  \fontdimen4\font\relax}
\providecommand{\BIBforeignlanguage}[2]{{%
\expandafter\ifx\csname l@#1\endcsname\relax
\typeout{** WARNING: IEEEtran.bst: No hyphenation pattern has been}%
\typeout{** loaded for the language `#1'. Using the pattern for}%
\typeout{** the default language instead.}%
\else
\language=\csname l@#1\endcsname
\fi
#2}}
\providecommand{\BIBdecl}{\relax}
\BIBdecl

\bibitem{yang2024sweagent}
J.~Yang, C.~E. Jimenez, A.~Wettig, K.~Lieret, S.~Yao, K.~Narasimhan, and
  O.~Press, ``{SWE}-agent: Agent-computer interfaces enable automated software
  engineering,'' in \emph{Advances in Neural Information Processing Systems
  (NeurIPS)}, 2024.

\bibitem{jimenez2024swebench}
C.~E. Jimenez, J.~Yang, A.~Wettig, S.~Yao, K.~Pei, O.~Press, and K.~Narasimhan,
  ``{SWE}-bench: Can language models resolve real-world {GitHub} issues?'' in
  \emph{The Twelfth International Conference on Learning Representations
  (ICLR)}, 2024.

\bibitem{robbes2026agentic}
R.~Robbes, T.~Matricon, T.~Degueule, A.~Hora, and S.~Zacchiroli, ``Agentic
  much? {A}doption of coding agents on {GitHub},'' \emph{ACM Transactions on
  Software Engineering and Methodology}, 2026.

\bibitem{li2026aidev}
H.~Li, H.~Zhang, and A.~E. Hassan, ``{AIDev}: Studying {AI} coding agents on
  {GitHub},'' in \emph{Proceedings of the 23rd International Conference on
  Mining Software Repositories (MSR)}, 2026.

\bibitem{robles2027wanted}
G.~Robles, D.~M. German, and B.~Vasilescu, ``I wanted to be a developer... and
  ended up a manager,'' 2027, submitted to the 20th International Conference on
  Cooperative and Human Aspects of Software Engineering (CHASE 2027). Under
  review.

\bibitem{mozannar2024reading}
H.~Mozannar, G.~Bansal, A.~Fourney, and E.~Horvitz, ``Reading between the
  lines: Modeling user behavior and costs in {AI}-assisted programming,'' in
  \emph{Proceedings of the CHI Conference on Human Factors in Computing Systems
  (CHI)}, 2024, pp. 1--16.

\bibitem{liang2024survey}
J.~T. Liang, C.~Yang, and B.~A. Myers, ``A large-scale survey on the usability
  of {AI} programming assistants: Successes and challenges,'' in
  \emph{Proceedings of the 46th IEEE/ACM International Conference on Software
  Engineering (ICSE)}, 2024, pp. 1--13.

\bibitem{becker2025metr}
J.~Becker, N.~Rush, B.~Barnes, and D.~Rein, ``Measuring the impact of
  early-2025 {AI} on experienced open-source developer productivity,''
  arXiv:2507.09089, 2025.

\bibitem{kosmyna2025brain}
N.~Kosmyna, E.~Hauptmann, Y.~T. Yuan, J.~Situ, X.-H. Liao, A.~V. Beresnitzky,
  I.~Braunstein, and P.~Maes, ``Your brain on {ChatGPT}: Accumulation of
  cognitive debt when using an {AI} assistant for essay writing task,''
  arXiv:2506.08872, 2025.

\bibitem{risko2016offloading}
E.~F. Risko and S.~J. Gilbert, ``Cognitive offloading,'' \emph{Trends in
  Cognitive Sciences}, vol.~20, no.~9, pp. 676--688, 2016.

\bibitem{kazemitabaar2023novices}
M.~Kazemitabaar, J.~Chow, C.~K.~T. Ma, B.~J. Ericson, D.~Weintrop, and
  T.~Grossman, ``Studying the effect of {AI} code generators on supporting
  novice learners in introductory programming,'' in \emph{Proceedings of the
  CHI Conference on Human Factors in Computing Systems (CHI)}, 2023, pp. 1--23.

\bibitem{prather2023robots}
J.~Prather, P.~Denny, J.~Leinonen, B.~A. Becker \emph{et~al.}, ``The robots are
  here: Navigating the generative {AI} revolution in computing education,'' in
  \emph{Proceedings of the 2023 Working Group Reports on Innovation and
  Technology in Computer Science Education (ITiCSE-WGR)}, 2023, pp. 108--159.

\bibitem{shen2026skill}
J.~H. Shen and A.~Tamkin, ``How {AI} impacts skill formation,''
  arXiv:2601.20245, 2026.

\bibitem{nagappan2005churn}
N.~Nagappan and T.~Ball, ``Use of relative code churn measures to predict
  system defect density,'' in \emph{Proceedings of the 27th International
  Conference on Software Engineering (ICSE)}, 2005, pp. 284--292.

\bibitem{bird2011ownership}
C.~Bird, N.~Nagappan, B.~Murphy, H.~Gall, and P.~Devanbu, ``Don't touch my
  code! {E}xamining the effects of ownership on software quality,'' in
  \emph{Proceedings of the 19th ACM SIGSOFT Symposium and the 13th European
  Conference on Foundations of Software Engineering (ESEC/FSE)}, 2011, pp.
  4--14.

\bibitem{jetbrains2026aianalytics}
{JetBrains}, ``{AI} activity and impact,'' JetBrains Console documentation,
  \url{https://www.jetbrains.com/help/air/ai-activity-and-impact.html}, 2026,
  accessed 2026-10-25.

\bibitem{gitai2026}
{Git AI}, ``Git {AI}: Tracking {AI}-generated code in {Git},''
  \url{https://usegitai.com/}, 2026, accessed 2026-10-25.

\bibitem{bacchelli2013review}
A.~Bacchelli and C.~Bird, ``Expectations, outcomes, and challenges of modern
  code review,'' in \emph{Proceedings of the 35th International Conference on
  Software Engineering (ICSE)}, 2013, pp. 712--721.

\bibitem{mockus2002expertise}
A.~Mockus and J.~D. Herbsleb, ``Expertise browser: A quantitative approach to
  identifying expertise,'' in \emph{Proceedings of the 24th International
  Conference on Software Engineering (ICSE)}, 2002, pp. 503--512.

\bibitem{fritz2010dok}
T.~Fritz, J.~Ou, G.~C. Murphy, and E.~Murphy-Hill, ``A degree-of-knowledge
  model to capture source code familiarity,'' in \emph{Proceedings of the 32nd
  ACM/IEEE International Conference on Software Engineering (ICSE)}, 2010, pp.
  385--394.

\bibitem{fritz2007activity}
T.~Fritz, G.~C. Murphy, and E.~Hill, ``Does a programmer's activity indicate
  knowledge of code?'' in \emph{Proceedings of the 6th Joint Meeting of the
  European Software Engineering Conference and the ACM SIGSOFT Symposium on the
  Foundations of Software Engineering (ESEC/FSE)}, 2007, pp. 341--350.

\bibitem{bjork1994desirable}
R.~A. Bjork, ``Memory and metamemory considerations in the training of human
  beings,'' in \emph{Metacognition: Knowing about Knowing}, J.~Metcalfe and
  A.~P. Shimamura, Eds.\hskip 1em plus 0.5em minus 0.4em\relax MIT Press, 1994,
  pp. 185--205.

\bibitem{perry2023insecure}
N.~Perry, M.~Srivastava, D.~Kumar, and D.~Boneh, ``Do users write more insecure
  code with {AI} assistants?'' in \emph{Proceedings of the ACM SIGSAC
  Conference on Computer and Communications Security (CCS)}, 2023, pp.
  2785--2799.

\bibitem{kononenko2015review}
O.~Kononenko, O.~Baysal, L.~Guerrouj, Y.~Cao, and M.~W. Godfrey,
  ``Investigating code review quality: Do people and participation matter?'' in
  \emph{Proceedings of the IEEE International Conference on Software
  Maintenance and Evolution (ICSME)}, 2015, pp. 111--120.

\bibitem{fregnan2022first}
E.~Fregnan, L.~Braz, M.~D'Ambros, G.~Calikli, and A.~Bacchelli, ``First come
  first served: The impact of file position on code review,'' in
  \emph{Proceedings of the 30th ACM Joint European Software Engineering
  Conference and Symposium on the Foundations of Software Engineering
  (ESEC/FSE)}, 2022, pp. 483--494.

\bibitem{tornhill2015crime}
A.~Tornhill, \emph{Your Code as a Crime Scene}.\hskip 1em plus 0.5em minus
  0.4em\relax The Pragmatic Bookshelf, 2015.

\bibitem{hassan2009entropy}
A.~E. Hassan, ``Predicting faults using the complexity of code changes,'' in
  \emph{Proceedings of the 31st International Conference on Software
  Engineering (ICSE)}, 2009, pp. 78--88.

\bibitem{vaithilingam2022expectation}
P.~Vaithilingam, T.~Zhang, and E.~L. Glassman, ``Expectation vs. experience:
  Evaluating the usability of code generation tools powered by large language
  models,'' in \emph{CHI Conference on Human Factors in Computing Systems
  Extended Abstracts (CHI EA)}, 2022, pp. 1--7.

\bibitem{barke2023grounded}
S.~Barke, M.~B. James, and N.~Polikarpova, ``Grounded copilot: How programmers
  interact with code-generating models,'' \emph{Proceedings of the ACM on
  Programming Languages}, vol.~7, no. OOPSLA1, pp. 85--111, 2023.

\bibitem{watanabe2025agentic}
M.~Watanabe, H.~Li, Y.~Kashiwa, B.~Reid, H.~Iida, and A.~E. Hassan, ``On the
  use of agentic coding: An empirical study of pull requests on {GitHub},''
  \emph{ACM Transactions on Software Engineering and Methodology}, 2025,
  accepted. arXiv:2509.14745.

\bibitem{lee2025critical}
H.-P. Lee, A.~Sarkar, L.~Tankelevitch, I.~Drosos, S.~Rintel, R.~Banks, and
  N.~Wilson, ``The impact of generative {AI} on critical thinking:
  Self-reported reductions in cognitive effort and confidence effects from a
  survey of knowledge workers,'' in \emph{Proceedings of the CHI Conference on
  Human Factors in Computing Systems (CHI)}, 2025.

\bibitem{brooks1983comprehension}
R.~Brooks, ``Towards a theory of the comprehension of computer programs,''
  \emph{International Journal of Man-Machine Studies}, vol.~18, no.~6, pp.
  543--554, 1983.

\end{thebibliography}

\end{document}